\documentclass[%
reprint,
showpacs,
showkeys,
amsmath,amssymb,
aps,
prd,
]{revtex4-1} 

\usepackage{graphicx}
\usepackage{pgffor}
\usepackage{capt-of}
\usepackage{soul}
\usepackage{lipsum}

\newcommand{\uvozovky}[1]{``#1''}

\newcommand{\zav}[1]{\left( #1 \right)}
\def\beq{\begin{equation}}
\def\eeq{\end{equation}}
\def\bea{\begin{eqnarray}}
\def\eea{\end{eqnarray}}

\let\phi=\varphi

\let\phi=\varphi
\let\rho=\varrho

\begin{document}

\frenchspacing

\author{Martin Blaschke}
\email{martin.blaschke@fpf.slu.cz}
\author{Zden\v{e}k Stuchl\'{\i}k}
\email{zdenek.stuchlik@fpf.slu.cz}
\affiliation{%
Institute of Physics and Research Centre of Theoretical Physics and Astrophysics, Faculty of Philosophy and Science,
Silesian University in Opava,\\
Bezru\v{c}ovo n\'am\v est\'i~13, CZ-746\,01 Opava, Czech Republic}
\author{Filip Blaschke}
\email{filip.blaschke@fpf.slu.cz}
\affiliation{%
Institute of Physics and Research Centre of Theoretical Physics and Astrophysics, Faculty of Philosophy and Science,
Silesian University in Opava,\\
Bezru\v{c}ovo n\'am\v est\'i~13, CZ-746\,01 Opava, Czech Republic\\
and\\
Institute of Experimental and Applied Physics, Czech Technical University in Prague, Horsk\'a 3a/22, 128 00 Praha 2, Czech Republic}
\author{Petr Blaschke}
\email{petr.blaschke@math.slu.cz}
\affiliation{%
Mathematical institute in Opava, Silesian University in Opava,\\
Na Rybn\'i\v{c}ku 626/1, CZ-746\,01 Opava, Czech Republic
}

\title[] {Classical corrections to black hole entropy in $d$ dimensions: a rear window to quantum gravity?}

\begin{abstract}

We provide a simple derivation of the corrections for
Schwarzschild and Schwarzschild-Tangherlini black hole entropy without knowing the details of quantum gravity. We will follow Bekenstein, Wheeler and Jaynes ideas, using summations techniques without calculus approximations, to directly find logarithmic corrections to well-known entropy formula for black holes. Our approach is free from pathological behaviour giving negative entropy for small values of black hole mass $M$. With the aid of \uvozovky{Universality} principle we will argue that this purely classical approach could open a window for exploring properties of quantum gravity.   

\end{abstract}


\maketitle 
\section{Introduction}

It is well-known that there are different theoretical methods for derivation
of entropy $S$ for the black hole (BH) such as in string theories, loop quantum gravity, conformal symmetry near horizon, etc. \cite{Pan:Cig:Iva:2008:}. Surprisingly, all of these
methods give essentially the same result. This is an example of \uvozovky{Universality} principle which suggests that some underlying classical feature of the theory may control quantum states \cite{Cap:2009:}.   

Understanding BH entropy
is a challenge in modern physics. Any developments in this direction might lead to important insights into the structure of quantum gravity which includes in particular the notion of \uvozovky{holography} and the
emerging notion of \uvozovky{quantum spacetime}.

As noted by Bekenstein \cite{Bek:1998:}, BH is like a hydrogen atom in the field of the strong gravity regime.

Semiclassical quantization of black holes  has been calculated by several authors (\cite{Fra:2005:},\cite{Nic:Pee:Zam:2005:}) by making use of the
Hamilton-Jacobi formalism. For the stable circular orbits they find the
formula
\begin{equation}
E_n \sim M- \frac{2 G^2 M^5}{n^2 \hbar^2}\, 
\end{equation}
which is Bohr's formula for the Coulomb
interaction.

Thermodynamical characteristics of a BH represent one of the most important
subjects of the contemporary physics \cite{Sam:Cho:2007:} since Bekenstein \cite{Bek:1973:} suggested that a BH contains the entropy $S$
proportional to the horizon surface area, $A$. For the Schwarzschild's black
hole one has:
\begin{equation}\label{BHE}
S= \frac{k_{\mathrm{B}} c^3}{4 \hbar G} A
\end{equation}
where $k_{\mathrm{B}}$ is Boltzmann's constant, $c$ speed of light and $\hbar$ reduced Planck
constant. Also, Bekenstein suggested that the horizon surface area is
quantized, and can be changed only discretely. Bekenstein's analysis is, on the
one hand, based on the characteristics of corresponding, complex quantum
measurement procedures, i.e. Heisenberg's uncertainty relations and
Ehrenfest's adiabatic theorem. On the other hand, it relies on general
relativistic and quantum field theoretical requirement on the stability of
the capture of a quantum system within black hole. According to this
requirement, roughly speaking, Compton wavelength of a given quantum
system must be smaller than double of the Schwarzschilds radius, otherwise a
quantum system might escape from black hole by means of quantum tunneling \cite{Pag:2005:}. 

The simple, structureless picture of BHs changed dramatically with Hawking's discovery \cite{Haw:1971:} that the area of the BH event horizon cannot decrease in any physical process and by the discovery of four \textit{Laws of Black Hole Mechanics} \cite{Bar:Car:Haw:1973:}.

Bekenstein \cite{Bek:1973:} was the first to go where no man has gone before and boldly propose that black holes actually possess entropy. He asserted that the horizon area was a measure of how much entropy a black hole
could have, in sharp contrast to standard thermodynamic notions where entropy is supposed to be a function of volume. In modern view on entropy \cite{Jay:1982:}, this can be interpreted as having  less than complete information about the system,
and this hidden information manifests as entropy. In case of black holes, this hidden  information is contained in our
lack of information about the nature of gravitational collapse. Thus, the parameters characterizing BHs in GR actually do not specify individual BHs; rather they, like the temperature and pressure of gas, are mean parameters describing equivalence classes of BHs, each of which may have collapsed from a different star via different process.

With the semiclassical approximation, the BH entropy obeys the Bekenstein-Hawking
area law. When full quantum effects are considered, the area
law should undergo corrections, and these corrections
can be obtained from field theory methods \cite{Fur:1995:,Man:Sol:1998:,Pag:2005:}, quantum geometry techniques \cite{Rom:Par:2000:,Klo:Bra:DeB:2008:}, general statistical mechanical arguments \cite{Das:Kau:Maj:2001:,Che:Li:sha:2011:}, Cardy formula \cite{Car:2000:} and others. All these approaches show that the leading correction is logarithmic:
\begin{equation}\label{entc}
S= S_0 +\alpha \ln S_0 + \dots\, ,
\end{equation}
where $\alpha$ is a dimensionless constant and $S_0$ denotes the
uncorrected semiclassical entropy of a BH measured in bits.

It is well known that the corrected entropy formula of Eq. (\ref{entc}) strongly suggest the Universality. 

The investigation of the corrected entropy formula of high dimensional black holes in tunneling perspective was done in \citep{Zhu:Ren:Li:2009:}. 

In this work we shall determine, in a simple way, the thermodynamical characteristics of a Schwarzschild black hole: namely Bekenstein-Hawking entropy. We will demonstrate that Bekenstein original idea, taken in a solemn and considered manner could shed some light on the properties of quantum gravity.

\section{It from bit}

We are going to \uvozovky{build up} a Schwarzschild BH of mass $M$ and count its entropy $S$ by repeatedly sending quantum states with only one bit of information under the event horizon. Here by \uvozovky{quantum states} we mean in general any matter which carries specific amount of information.
 
Similar procedure was used, e.g. in \cite{Pan:Pre:Gru:2008:}, where the authors used such quantum states that circumference of a great
circle at BH horizon contained integer number of corresponding reduced Compton wavelength.  
This approach is analogous to the Bohr quantization postulate, according to which circumference of an electron circular orbit comprises an integer number of corresponding de Broglie's
wavelengths.

We will not use circumference of a black hole to determine properties of the quantum state, because this quantity is not sufficiently general. Rather, we shall use the surface area $A$ multiplied by some parameter $k_\mathrm{c}$ determining a cross section of a BH. We will argue that this parameter $k_\mathrm{c}$ could be actually the only thing in the formula for BH entropy, which can be potentially influenced by, so far unknown, underlying quantum gravity nature of the universe. By this we mean that the Bekenstein approach to BH entropy might be correct even for microscopic BHs.   

The main tool that we will use to obtain our goal will be summation operator. We will not use calculus approximation in any step, i.e. the BH mass $M$ cannot be changed smoothly, but it is also \uvozovky{quantized}.   

To count entropy we will follow Bekenstein idea. We will view upon an entropy as an amount of \textit{hidden}, or more appropriately in case of black holes, \textit{lost} information \cite{Jay:1982:} (although there are many who would argue, that no information can be lost, see e.g. \citep{Sus:2013:}). 

The following analysis will be done in arbitrary $d+1$ spacetime. The reason why we concentrate on general case is to directly support universality of BH entropy. This principle will be important in subsequent discussions.
    
Let us consider a \uvozovky{photon} falling into a black hole, carrying only one bit of information. In order to posses only one bit a photon should have wavelength: 
\begin{equation}\label{lambda}
\lambda =  k_\mathrm{c} \left(\frac{A \Gamma \left(\frac{d}{2}\right)}{2 \pi^{\frac{d}{2}}}\right)^{\frac{1}{d-1}}\, ,
\end{equation}  
where $d$ is the spacial dimension of the spacetime and $k_{\mathrm{c}}$ is a constant related to cross section of this process. 

In case of $d=3$ the previous equation reduces to
\begin{equation}
\lambda =  k_\mathrm{c} \sqrt{\frac{A}{4\pi}}\, .
\end{equation}
This case illustrates the idea clearly. Wavelength is chosen to be of the same order as the area radius of the black hole.
Photon with this  wavelength, when it is under the horizon, \uvozovky{hides} one bit of information, because it can give answer to only one question: \text{do you exist?} 
Obviously, here we neglect any polarization or other quantum numbers. We use photons, because they are intuitively very appropriate to demonstrate this idea following Wheeler's \uvozovky{it from bit} doctrine.

To be more clear, we could use other kinds of particles and other processes of adding one bit of information into a black hole. However, if we assume that information is a scalar entity and there are no different kinds of information in the nature, all these processes should lead to the same answer.  

The photon has to have this particular wavelength, so that we are sure that we add only one bit of information into the BH. To make this notion clearer, let us imagine that the photon has larger wavelength. In this case, we can have scattering revealing information about size of the horizon, and we are no longer sure, how much information fell into the BH. On the other hand, photon with shorter wavelengths would surely fall into the BH, but we would be able to ask for more concrete information, e.g. where the photon entered the horizon, and so on. Only with the specific wavelength Eq. (\ref{lambda}) we can only be sure about one thing, that the photon is inside. So this particular wavelength represents minimal informational gain for BH. 

Because, we oversimplify the situation, we ignore the polarization and claim that this minimum of information is 1 bit. In practice, this minimum would be easier to achieve by some hypothetical massless, spineless scalar particle.   

The processes such as tunneling and Hawking radiation can introduce some technical difficulties (for instance, precise determination of the scattering constant $k_\mathrm{c}$), but they change nothing in principle, if we believe that Hawking radiation carries out the original bits of information, (to further clarify this argument, we refer to articles about black hole information paradox).   

Photon with wavelength $\lambda$ increases BH's mass by:
\begin{equation}\label{delta}
\Delta M = \frac{h}{\lambda c}\, ,
\end{equation} 
where $h$ is Planck constant. Meanwhile, the amount of hidden information, increases by one bit, which is equivalent with increase in entropy by one Boltzmann constant: $\Delta S=1 k_{\mathrm{B}}\,$.
Because the area of a BH depends on its mass, we can write down a difference equation.

The area of a Schwarzschild-Tangherlini black hole \cite{Tan:1963:}, a direct generalization of Schwarzschild black hole to higher dimensions, reads
\begin{equation}\label{area}
A = \frac{2 \pi^\frac{d}{2}}{\Gamma\left(\frac{d}{2}\right)}\left(\frac{16\pi G_{(d+1)} M \Gamma\left(\frac{d}{2}\right)}{c^2 (d-1) 2 \pi^\frac{d}{2}}\right)^{\frac{d-1}{d-2}}\, ,
\end{equation} 
where $G_{(d+1)}$ is Newton's gravitational constant in $d+1$ dimensions.  
 
Putting (\ref{lambda}) with the use of (\ref{area}) into (\ref{delta}) we find that the difference equation is
\begin{equation}\label{de}
\displaystyle \Delta M = K_d M^{\frac{1}{2-d}}\, ,
\end{equation}
where we have done the substitution
\begin{equation}
K_d=\frac{h}{c k_\mathrm{c}}\left(\frac{8 G_{d+1} \Gamma\left(\frac{d}{2}\right)}{c^2 (d-1)\pi^\frac{d-2}{2}}\right)^{\frac{1}{2-d}}\, .
\end{equation}

At this point, the difference equation (\ref{de}) is usually approximated by corresponding differential equation. This is well justified for \uvozovky{end part} of the summation, where $M$ of a black hole can be macroscopically
large and so 
\begin{equation}
\frac{\Delta S}{\Delta M} \sim \frac{\mathrm{d} S}{\mathrm{d M}}\, , \quad M\rightarrow \infty\, .
\end{equation}
Using this approximation we would find usual Bekenstein-Hawking entropy law (\ref{BHE}). However, we will not use this approximation because it is especially not suitable in the \uvozovky{beginning} part of a summation, where $\Delta M$ can be comparable with respect to $M$.    

Applying the summation operator $\sum$ (summation equivalent to indefinite integral operator) to Eq. (\ref{de}) we obtain mass $M$ as a function of entropy $S$. 
\begin{equation}\label{er}
\frac{S}{k_{\mathrm{B}}} = \sum\frac{\Delta M}{K_{d}M^\frac{1}{2-d}}\, .
\end{equation}

\subsection{Case $d=3$ -- Schwarzschild geometry}

In case of (3+1)-dimensional universe, the difference equation has the form:
\begin{equation}\label{dee3}
M \Delta M = K_3\, .
\end{equation}
Using \uvozovky{per partes} method \cite{Boo:1880:} we have 
\begin{equation}\label{perpar}
\sum M(n) \Delta M(n) = M^2(n) - \sum M(n+1) \Delta M(n)\, . 
\end{equation}
Here we use $n$ as a summation variable. Furthermore, if we assume asymptotic relation
\begin{equation}\label{ass}
\sum M(n+1)\Delta M(n) \sim \sum M(n)\Delta M(n)\, , 
\end{equation}
as $M$ goes to  $\infty$, we discover 
\begin{equation}
\sum M\Delta M \sim \frac{M^2}{2}\, , \quad M\rightarrow\infty\, 
\end{equation} 
and the solution to Eq. (\ref{de}) reads
\begin{equation}\label{RHS}
\sum\limits_{n=1}^{N} M\Delta M \sim \frac{M^2}{2} - \frac{M^2(1)}{2} = N K_3\, .
\end{equation}
Here $N$ is a final number of photons carrying one bit of information thrown into a BH, and $M(1)$ represents the mass of the BH with one bit of hidden information in it. This represents a cut off to the theory, which naturally arise from the fact that we are using summation techniques instead of continuous calculus. In other words, the summation techniques force us to assume that there is a minimal mass of a BH. This also suggests that there is a maximal mass for a photon carrying one bit of information. This maximally massive state can be envisioned as a state of a photon in vacuum just before collapsing on its own mass into smallest possible black hole. 

In order to justify RHS of Eq. (\ref{RHS}), we suppose that such photon caries 1 bit of information, but of course, it can be otherwise. It could be that smallest existing black hole have more than one bit of hidden information. But our setup is more natural.  

We can summarized this into one assumptions:
\\
\\
\textit{There is a maximal possible size/mass for a quantum state (photon) with 1 bit of information.}    
\\
\\
The relation between $N$ and entropy $S$ is then
\begin{equation}
\frac{S}{k_\mathrm{B}} = N \Rightarrow S \sim k_{\mathrm{B}}\frac{M^2-M^2(1)}{K_3}\, , \quad M\rightarrow\infty\, .
\end{equation}
If we ignore entropy of the smallest possible BH $(M(1)\ll  M)$ we see that we have rediscovered (with suitable choice of constant $k_\mathrm{c}=4\pi^2$) the Bekenstein-Hawking area law for entropy (\ref{BHE}). 

Let us find the next term in asymptotic expansion of entropy. From Eq. (\ref{dee3}) we express 
\begin{equation}
M(n+1) = M(n) + \frac{K_3}{M(n)}\, .
\end{equation} 
Putting this result into (\ref{perpar}) we have
\begin{equation}
2\sum M(n) \Delta M(n) = M^2(n) - K_3\sum  \frac{\Delta M(n)}{M(n)}\, . 
\end{equation}
Now, if we break our promise and do calculus approximation on the last sum in this expression, we immediately get logarithmic correction via asymptotic relation: 
\begin{equation}
\sum \frac{\Delta M(n)}{M(n)} \sim \int \frac{\mathrm{d} M(n)}{M(n)} = \ln\left[\frac{M(n)}{M(1)}\right] + c \, ,
\end{equation} 
as $M$ goes to  $\infty$ \footnote{Note that this \uvozovky{naive} approximation is not always correct. For example, function $M(n)=2^n$ is growing so fast that the $\Delta 2^n$ which is equal to $2^n$, cannot be approximated by its differential $\mathrm{d} 2^n/\mathrm{d} n$ for any value of $n$.}. Here $c$ is an integration constant and $M(1)$ was put into logarithm to assure dimensionlessness of the argument. 

Let us stress out that we do not have to use calculus approximation in any degree of expansion. See Appendix A for exact asymptotic expansion for the solution of difference equation (\ref{dee3}).

Putting all relations together, we have
\begin{equation}\label{ro}
\frac{M^2}{2}-\frac{K_3\ln \left(X^2\right)}{4}\sim K_3 N\, ,\quad M\rightarrow\infty\, ,
\end{equation}
where we have used substitution
\begin{equation}
\frac{M}{M(1)}\equiv X
\end{equation}

\begin{figure}[t]
\begin{center}
\centering
\includegraphics[width=1\linewidth]{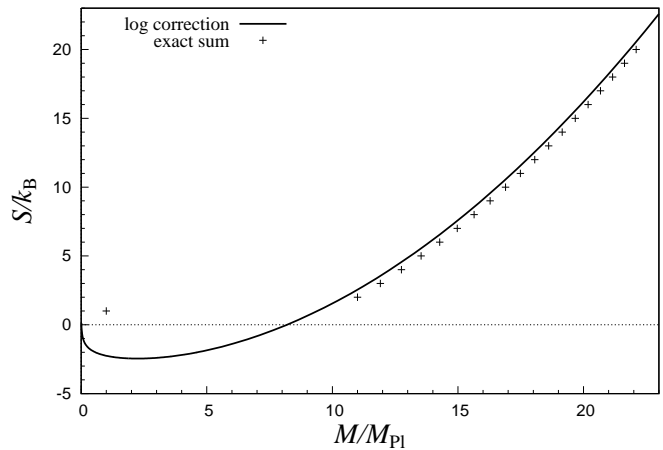}
\caption{\label{ff} Illustrative example, $M(1)=M_{\mathrm{Pl}}$ and  $K_3=10 M^2_{\mathrm{Pl}}$. We can see that uncorrected, exact entropy calculated by summation (\ref{er}), denoted by $+$, do not have negative entropy problems. The full line represents continuous approximation with logarithmic correction. Note, that in case of Bekenstein-Hawking area law, the $K_3=1/(16 \pi) M^2_{\mathrm{Pl}}$ . We use an unusual value $K_3=10 M^2_{\mathrm{Pl}}$ to emphasize unphysical effect of decreasing entropy for BHs with small masses.}
\end{center}
\end{figure}

Putting $N=1$ into Eq. (\ref{ro}) we see that the minimal mass of BH with 1 bit of entropy is $M(1)=\sqrt{2K_3}\, .$

The entropy formula is thus:
\begin{equation}\label{fr}
S \sim \frac{k_{\mathrm{B}}}{K_3}\left[\frac{M^2}{2}-\frac{K_3 \ln\left(X^2\right)}{4}\right]\, ,\quad M\rightarrow\infty\, .
\end{equation} 
The complete expansion of the Schwarzschild BH entropy is presented in Apendix A.

We have rediscovered logarithmic corrections to Bekenstein--Hawking entropy (\ref{entc}).    
In Fig. \ref{ff} we can see one of the main results of this article. The entropy computed by summation techniques is not demonstrating pathological behaviour for small values $M$ as occurs for the continuous approximation. Namely, the entropy is not negative and it is never decreasing function of $M$. 
This suggests that original Bekenstein and Hawking approach to BH entropy, if taken seriously, seems correct even for the smallest possible BHs. It is possible that entropy calculated in this way is quantitatively correct and any corrections coming from quantum gravity might just influence the cross section constant $k_{\mathrm{c}}$. This logic strongly depends on our understanding of the relation between information and entropy.      

\subsection{Case with arbitrary $d$ -- Schwarzschild--Tangherlini geometry}

Here we could repeat the same procedure as in case 3+1. Starting with difference equation on general dimension

\begin{equation}
M^{\frac{1}{d-2}}\Delta M = K_d\, ,
\end{equation}
we would rediscover Bekenstein--Hawking area law as a lowers order in asymptotic expansion. Finding the next terms is quite similar to the previous case. The only new information we need is 
\begin{equation}
\Delta M^d \sim d M^{d-1} \Delta M\, ,\quad M\rightarrow \infty\, .
\end{equation}  
Using this we find that the next term in asymptotic expansion is:
\begin{equation}
-\frac{1}{2\left(d-1\right)}\ln\left(X^2\right) \, 
\end{equation}
and do not depend on the constant $K_d$. 
This result strongly supports the universality principle for BH entropy. 

The universality principle is essential element to our argument. As we have mentioned in the introduction, the leading correction to area law of BH entropy is very robust entity. It is independent of the underlying approach and has always logarithmic form. In fact, for $D=4$ spacetimes, it can be shown that principle of maximum entropy leads to Einstein equations as equations of state for spacetime \cite{Jac:1995:}, and for $D>4$ to Lanczos-Lovelock models \cite{Pad:2010:}. This system behaves similarly like critical point in phase transition and therefore throws away all unimportant informations. Therefore, if handled correctly, it could represent a window to quantum gravity in the same manner as critical points represent windows from thermodynamic description of gases to statistical and kinetic theories of particles.

\section{Conclusions}

We have shown that by following classical Bekeinstein idea, supported by modern view on entropy (Jaynes) and Wheeler's \textit{It from bit} doctrine, we can calculate the same correction formulas for BH entropy. This result  supports universality principle and we argue that it opens a small window for quantum gravity. Summation techniques for obtaining BH entropy do not posses pathological behavior for small $M$, namely negative values of entropy and violation of second law of BH thermodynamics. It is quite interesting the \uvozovky{quantum} discrete calculus could solve this fundamental disease of the BH entropy calculation.  
Our work also indicate, that the correct version of quantum gravity should have a maximal mass for a state with one bit of information. 
In the future it would be interesting to find out how bit-by-bit doctrine would work in the theories which do not follow area law of entropy.         


\section{Appendix: Asymptotic expansion for $d=3$ case}

We are looking for asymptotic expansion for the solution of difference equation:
\begin{equation}\label{deA}
M\Delta M=K_3\, ,
\end{equation}
where the $\Delta$ operator is defined by relation:
\begin{equation}
\Delta M(n) = M(n+1) - M(n)\, .
\end{equation}

We know from the previous discussion that the solution to (\ref{deA}) is in the form
\begin{equation}\label{rovb}
n=\frac{M^2}{2K_3} -\frac{1}{4} \ln \left(X^2\right) + F(M^2)\, ,
\end{equation}
where 
\begin{equation}
F(M^2):=\frac14\ln \left(X^2\right)-\frac12\sum\frac{\Delta M}{M} ,
\end{equation}
is the error term and $X\equiv M/M(1)\, $. 

Applying the operator $\Delta$ on this, we see that
\[
2\Delta F=\ln\zav{1+\frac{\Delta M}{M}}-\frac{\Delta M}{M} \sim \frac12\zav{\frac{\Delta M}{M}}^2=\frac{K_3\Delta M}{2M^3}. 
\]
Hence 
\[
F=\frac{K_3}{4}\sum\frac{\Delta M}{M^3} = \frac{K_3}{4}\int \frac{{\rm d}M}{M^3}+\tilde F(M^2)=c-\frac{K_3}{8 M^2}+\tilde F\, ,
\]
where $c$ is summation constant and $\tilde F$ an another unknown function. 
Repeating the process we can see that the error term $F$ has an asymptotic expansion in negative even powers of $M$:
\begin{equation}
F(M^2):=c_0+\frac{c_1 K_3^2}{M^2}+\frac{c^2 K_3^3}{M^4}+\frac{c^3 K_3^4}{M^6}+\dots,
\end{equation}
for some constants $c_j$.

Substituting this expansion into (\ref{rovb}) and applying the $\Delta$ operator provides us with the asymptotic identity:
\begin{eqnarray}
2K_3&=&2\Delta M M+(\Delta M)^2-K_3\ln\zav{1+\frac{\Delta M}{M}}\nonumber\\
&+& \sum_{j=1}^{\infty}\frac{K_3^{j+1} c_j}{M^{2j}}\zav{\zav{1+\frac{\Delta M}{M}}^{-2j}-1}.\nonumber
\end{eqnarray}
Using the fact that $\Delta M=K_3/M$ and expanding all functions in negative powers of $M$ gives us
\[
0=K_3\sum_{j=2}^{\infty}\zav{-\frac{K_3}{M^2}}^j\frac{1}{j}+\sum_{j=1}^{\infty}\frac{K_3^{j+1}c_j}{M^{2j}}\sum_{\ell=1}^{\infty}\frac{(2j)_\ell}{\ell!}\zav{-\frac{K_3}{M^2}}^{\ell}.
\]
where $(j)_l$ is the Pochhammer symbol:
\begin{equation}
(j)_\ell \equiv \frac{\Gamma(j+\ell)}{\Gamma(j)}= j(j-1)...(j-\ell+1)\, .
\end{equation}
Collecting like powers of $M$ we obtain equations on the coefficients $c_j$ which gives us:
\begin{equation}
c_1=\frac{1}{4},\qquad c_2=\frac{5}{48},\qquad c_3=\frac76,\qquad \dots
\end{equation}

The resulting asymptotic expansion is in the form: 
\begin{equation}
n\sim \frac{M^2}{2K_3}-\frac{1}{4}\ln\left(X^2\right)+c_0+\frac{K_3 c_1}{2M^2}+\frac{K^2_3 c_2}{2M^4}+\dots\, ,
\end{equation}
as $M$ goes to  $\infty$.

\section*{Acknowledgements}

Authors would like to thank to Petr Slan\'y for useful discussions. 
MB, FB and ZS has been supported by the Albert Einstein Centre for Gravitation and Astrophysics financed by the Czech Science Agency Grant No. 14-37086G. 

FB is also supported by the program of Czech Ministry of Education Youth and Sports INTEREXCELLENCE Grant number LTT17018 and MB by the Silesian University at Opava grant SGS/14/2016. 

PB has been supported by GA \v CR grant no. 201/12/G028.

\section*{References}

\begin{thebibliography}{10}

\bibitem{Pan:Cig:Iva:2008:}
V.~{Pankovic}, S.~{Ciganovic}, and J.~{Ivanovic}.
\newblock {A Simple Determination of the (LOGARITHMIC) Corrections of Black
  Hole Entropy without Knowing the Details of Quantum Gravity}.
\newblock {\em ArXiv e-prints}, October 2008.

\bibitem{Cap:2009:}
S.~{Carlip}.
\newblock {\em {Black Hole Entropy and the Problem of Universality}}, page~91.
\newblock 2009.

\bibitem{Bek:1998:}
J.~D. {Bekenstein}.
\newblock {Black Holes: Classical Properties, Thermodynamics and Heuristic
  Quantization}.
\newblock {\em ArXiv General Relativity and Quantum Cosmology e-prints}, August
  1998.

\bibitem{Fra:2005:}
M.~{Frasca}.
\newblock {Existence of a semiclassical approximation in loop quantum gravity}.
\newblock {\em General Relativity and Gravitation}, 37:2239--2243, December
  2005.

\bibitem{Nic:Pee:Zam:2005:}
H.~{Nicolai}, K.~{Peeters}, and M.~{Zamaklar}.
\newblock {TOPICAL REVIEW: Loop quantum gravity: an outside view}.
\newblock {\em Classical and Quantum Gravity}, 22:R193--R247, October 2005.

\bibitem{Sam:Cho:2007:}
J.~{Samuel} and S.~R. {Chowdhury}.
\newblock {Geometric flows and black hole entropy}, June 2007.

\bibitem{Bek:1973:}
J.~D. {Bekenstein}.
\newblock {Black Holes and Entropy}.
\newblock {\em Physical Review D}, 7:2333--2346, April 1973.

\bibitem{Pag:2005:}
Don~N Page.
\newblock Hawking radiation and black hole thermodynamics.
\newblock {\em New Journal of Physics}, 7(1):203, 2005.

\bibitem{Haw:1971:}
S.~W. {Hawking}.
\newblock {Gravitational Radiation from Colliding Black Holes}.
\newblock {\em Physical Review Letters}, 26:1344--1346, May 1971.

\bibitem{Bar:Car:Haw:1973:}
J.~M. {Bardeen}, B.~{Carter}, and S.~W. {Hawking}.
\newblock {The four laws of black hole mechanics}.
\newblock {\em Communications in Mathematical Physics}, 31:161--170, June 1973.

\bibitem{Jay:1982:}
E.~T. {Jaynes}.
\newblock {On the Rationale of Maximum Entropy Methods}.
\newblock {\em IEEE Proceedings}, 70:939--952, 1982.

\bibitem{Fur:1995:}
Dmitri~V. Fursaev.
\newblock Temperature and entropy of a quantum black hole and conformal
  anomaly.
\newblock {\em Phys. Rev. D}, 51:R5352--R5355, May 1995.

\bibitem{Man:Sol:1998:}
Robert~B. Mann and Sergey~N. Solodukhin.
\newblock Universality of quantum entropy for extreme black holes.
\newblock {\em Nuclear Physics B}, 523(1):293 -- 307, 1998.

\bibitem{Rom:Par:2000:}
Romesh~K. Kaul and Parthasarathi Majumdar.
\newblock Logarithmic correction to the bekenstein-hawking entropy.
\newblock {\em Phys. Rev. Lett.}, 84:5255--5257, Jun 2000.

\bibitem{Klo:Bra:DeB:2008:}
S.~{Kloster}, J.~{Brannlund}, and A.~{DeBenedictis}.
\newblock {Phase space and black-hole entropy of higher genus horizons in loop
  quantum gravity}.
\newblock {\em Classical and Quantum Gravity}, 25(6):065008, March 2008.

\bibitem{Das:Kau:Maj:2001:}
S.~{Das}, R.~K. {Kaul}, and P.~{Majumdar}.
\newblock {New holographic entropy bound from quantum geometry}.
\newblock {\em Physical Review D}, 63(4):044019, February 2001.

\bibitem{Che:Li:sha:2011:}
Y.-X. {Chen}, J.-L. {Li}, and K.-N. {Shao}.
\newblock {Logarithmic corrections to black hole and black ring entropy in
  tunneling approach}.
\newblock {\em EPL (Europhysics Letters)}, 95:10008, July 2011.

\bibitem{Car:2000:}
S~Carlip.
\newblock Logarithmic corrections to black hole entropy, from the cardy
  formula.
\newblock {\em Classical and Quantum Gravity}, 17(20):4175, 2000.

\bibitem{Zhu:Ren:Li:2009:}
T.~{Zhu}, J.-R. {Ren}, and M.-F. {Li}.
\newblock {Corrected entropy of high dimensional black holes}.
\newblock {\em ArXiv e-prints}, June 2009.

\bibitem{Pan:Pre:Gru:2008:}
V.~{Pankovic}, M.~{Predojevic}, and P.~{Grujic}.
\newblock {Bohr's Semiclassical Model of the Black Hole Thermodynamics}.
\newblock {\em Serbian Astronomical Journal}, 176:15--22, June 2008.

\bibitem{Sus:2013:}
L.~{Susskind}.
\newblock {Black Hole Complementarity and the Harlow-Hayden Conjecture}.
\newblock {\em ArXiv e-prints}, January 2013.

\bibitem{Tan:1963:}
F.~R. {Tangherlini}.
\newblock {Schwarzschild Field in $n$ Dimensions and the Dimensionality of
  Space Problem}.
\newblock {\em Nuovo Cimento}, 27:636--651, 1963.

\bibitem{Boo:1880:}
G.~Boole and J.F. Moulton.
\newblock {\em A Treatise on the Calculus of Finite Differences}.
\newblock MacMillan and Company, 1880.

\bibitem{Jac:1995:}
T.~{Jacobson}.
\newblock {Thermodynamics of Spacetime: The Einstein Equation of State}.
\newblock {\em Physical Review Letters}, 75:1260--1263, August 1995.

\bibitem{Pad:2010:}
T.~{Padmanabhan}.
\newblock {Thermodynamical aspects of gravity: new insights}.
\newblock {\em Reports on Progress in Physics}, 73(4):046901, April 2010.

\end{thebibliography}

\end{document}